# Structural and Electrocatalytic Properties of La-Co-Ni Oxide Thin Films

Patrick Marx[1], Shivam Shukla[2], Alejandro Esteban Perez Mendoza[2], Florian Lourens[1], Corina Andronescu[2], Alfred Ludwig[1]

[1]Chair for Materials Discovery and Interfaces, Institute for Materials, Ruhr University Bochum, Universitätsstraße 150, 44801 Bochum, Germany

[2]Chemical Technology III, Faculty of Chemistry and CENIDE, Center for Nanointegration, University of Duisburg-Essen, Carl-Benz-Straße 199, 47057 Duisburg, Germany

*corresponding author: alfred.ludwig@rub.de

**Abstract:**
La-Co-Ni oxides were fabricated in the form of thin-film materials libraries by combinatorial reactive co-sputtering and analyzed for structural and functional properties over large compositional ranges: normalized to the metals of the film they span about 0 - 70 at.-% for Co, 18 - 81 at.-% for La and 11 - 25 at.-% for Ni. Composition-dependent phase analysis shows formation of three areas with different phase constitutions in dependance of Co-content: In the La-rich region with low Co content, a mixture of the phases $La_2O_3$, perovskite, and $La(OH)_3$ is observed. In the Co-rich region, perovskite and spinel phases form. Between the three-phase region and the Co-rich two-phase region, a single-phase perovskite region emerges. Surface microstructure analysis shows formation of additional crystallites on the surface in the two-phase area, which become more numerous with increasing Ni-content. Energy-dispersive X-ray analysis indicates that these crystallites mainly contain Co and Ni, so they could be spinels growing on the surface. The analysis of the oxygen evolution reaction (OER) electrocatalytic activity over all compositions and phase constitutions reveals that the perovskite/spinel two-phase region shows the highest catalytic activity, which increases with higher Ni-content. The highest OER current density was measured as 2.24 mA/cm$^2$ at 1.8 V vs. RHE for the composition $La_{11}Co_{20}Ni_9O_{60}$.

1. Introduction

Growing global awareness of environmental and climate protection drives further development of sustainable energy sources. The production of "green hydrogen", which is obtained from renewable energy sources such as solar or wind, plays a key role. Water splitting, powered by renewable energy, offers a way to produce high-purity sustainable hydrogen. To increase the efficiency of water splitting, high-performance electrocatalysts are required for the kinetically sluggish oxygen evolution reaction (OER). At the moment, precious metal-based catalysts are still mainly used, as they show the best catalytic activity and stability. For OER, $IrO_2$ and $RuO_2$ are benchmark catalysts due to their high intrinsic activity. [1,2] Due to their high cost and limited availability, much research has been done in recent years to replace them with more abundant materials. Various transition metals and alloys, as well as their oxides, nitrides, phosphides, sulfides, and also metal-organic frameworks have already been investigated as possible catalysts for water splitting. [3,4] Especially oxides crystallizing in the perovskite structure are recognized as promising catalysts for OER, sometimes rivaling or surpassing noble metal catalysts with respect to activity and stability. [5–8] Li et al. for example reported on $LaMnO_3$ nanosheets, which show at 10 mA cm$^{-2}$ a considerably smaller overpotential of 324 mV compared to $IrO_2$ with an overpotential of 379 mV. [7] An advantage of the perovskite structure ($ABO_3$) is its high compositional flexibility, which allows a wide variation of the elements for both the A and B sites. This compositional versatility enables tuning of structural and functional properties to optimize catalytic activity. [9,10] Due to this compositional tunability and the possibility for high catalytic activity and stability, La-based perovskites are a particular focus of research, with La occupying the A-sites and various transition metals being placed on the B-sites. [11–15]

The compositional possibilities offered by perovskites can be explored efficiently by high-throughput experimentation. Combinatorial magnetron sputtering of thin-film libraries offers the possibility to synthesize samples over large compositional ranges under the same conditions in the same experiment. So-called materials libraries (MLs) with a continuous chemical composition gradient (composition spread) are produced, which thus contain hundreds to thousands of different samples. High-throughput characterization methods are used to efficiently investigate the properties of the MLs, such as chemical composition, crystallographic structure, resistivity or catalytic activity. [16,17] Reactive combinatorial sputtering is suitable to fabricate and investigate La-based perovskites, because a self-organized growth of single-phase regions was observed, as shown in a previous work. [18] This is used to produce MLs which comprise single-phase perovskite regions which contain a wide variation of elements for the B-site atoms. [19] A low electrical resistance of the material can be beneficial for a high electrocatalytic performance. Therefore, Ni can be added to La-Co-perovskites, as it was reported to increase conductivity in several oxides. [20,21] Further, the addition of Ni was also reported to increase the catalytic activity of $LaCoO_3$ perovskites for OER. [22,23]

This work presents the combinatorial synthesis and high-throughput analysis of a La-Co-Ni-O ML, including the chemical composition, crystallographic structure, electrical resistivity and OER electrocatalytic activity. Furthermore, an analysis of the surface microstructure was carried out for selected measurement areas (MAs) and possible correlations of the different measurement values were examined.

## 2. Experimental Methods

*2.1 Synthesis of La-Co-Ni-O Thin-Film Library*

The investigated thin-film ML was synthesized by hot reactive co-sputtering in a commercial sputter system (AJA International, ATC 2200) with four confocally aligned cathodes. La was sputtered with a radio frequency power of 200 W from a $La_2O_3$ compound target (4-inch diameter, 99.99%, EvoChem). Co was sputtered with direct current power of 70 W from an elemental target (4-inch diameter, 99.99%, Sindlhauser Materials). Ni was sputtered with pulsed DC power of 70 W from an elemental target (4-inch diameter, 99.99%, EvoChem). The $La_2O_3$- and Co-targets were positioned on the cathodes 180° opposite to each other, and the Ni-target was positioned on a cathode oriented 90° from the others. A sketch of the sputter process and cathode positions is shown in Figure 1. The base pressure of the system was around $10^{-5}$ Pa. The substrate was heated to 500°C during the deposition using the in-built resistive heater. The substrate was kept at this temperature during the deposition and cooled down in vacuum afterwards. The deposition was performed in a reactive $Ar/O_2$ atmosphere with a constant $Ar/O_2$ flow ratio of 80 sccm/40 sccm at a fixed pressure of 0.4 Pa. The deposition was carried out for 10800 s on a 4-inch diameter sapphire wafer (c-plane) to obtain a thin-film thickness of 300 nm. For electrochemical measurements, the ML was deposited a second time on a platinized 4-inch (100)-Si-wafer (with 50 nm thermal $SiO_2$). The Pt-layer is needed as a bottom electrode for electrochemical measurements and has a thickness of around 50 nm. A 10 nm adhesion layer is underneath the Pt-layer to increase the adhesion of Pt on $SiO_2$. The deposition of the oxide ML was carried out for 2880 s to obtain a thin-film thickness of approximately 80 nm. This rather small thin-film thickness is needed to enable electrochemical measurements.

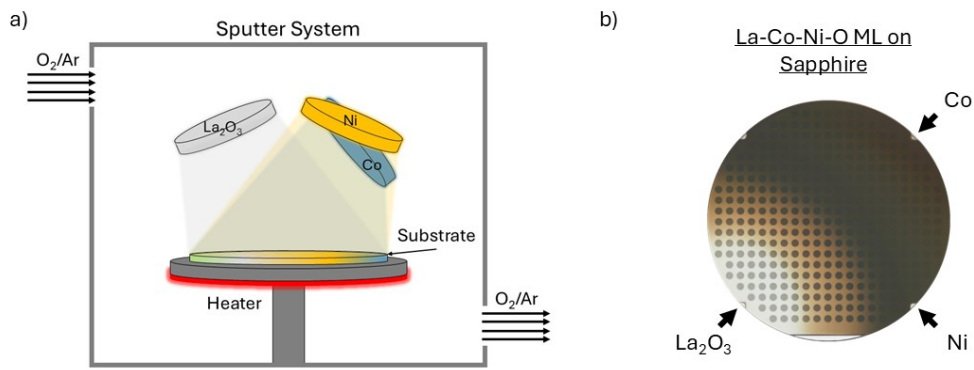

Fig. 1: a) Schematic of the reactive sputter co-deposition process of the La-Co-Ni-O ML. b) Photo of the ML deposited on a 100 mm diameter sapphire wafer with indicated target positions. The dots are from a representation of the MAs printed in the background of the photo and are visible through the (semi)transparent areas of the ML.

## 2.2 High-Throughput Characterization

For systematic characterization, a measurement scheme of 342 square-shaped MAs with a size of 4.5 mm x 4.5 mm was defined for the ML. This scheme was used for all automated high-throughput characterization methods to ensure their direct comparability.

The chemical composition with respect to the metals in the films on all MAs on the ML was automatically measured by EDX in a SEM (JEOL JSM-7200F with Oxford AZtecEnergy X-MaxN 80 $mm^2$ - SDD detector). The system was calibrated on a Cu-standard before the measurement. An acceleration voltage of 15 kV with a magnification of 600x was used for the SEM, whereby an area of 400 µm x 600 µm was measured within each MA. The EDX-data is normalized to the detected metals in the thin film excluding the O- and substrate signals.

X-ray Photoelectron Spectroscopy (XPS) was used to measure the O-content of the thin films in selected MAs. These MAs were selected after XRD phase constitution analysis to represent each identified phase area. The XPS measurements were carried out on a Kratos Axis Nova, using a monochromatic Al Kα X-ray source operating at 180 W (15 mA emission current, 12 kV anode voltage). The delay-line detector was set to pass energies of 160 eV for survey scans and 20 eV for narrow scans. The charge neutralizer was turned on, and there was no sputter etching applied. The pressure during the measurements was around 5*10$^{-7}$ Pa. At each of the selected MAs, the XPS analysis areas were 300 x 700 µm$^2$. Narrow scans were recorded of La 3d, Co 2p, Co 3p (including the overlapping Ni 3p), O 1s and C 1s. The latter was used for charge correction (adventitious C–C at 284.8 eV). The data was quantified with the Kratos ESCApe 1.4 software with its predefined relative sensitivity factors, considering the peaks of La 3d5/2, Co 2p3/2 (in two cases Co 2p1/2), Ni 3p and O 1s.

For phase analysis, X-ray diffraction (XRD) was performed on a Bruker D8 Discovery with 2D-detector (Vantec 500) and a Cu K$\alpha$ source (50 W, 0.15418 nm). An angular range for 2Θ of 15° to 75° was obtained for each MA. For peak analysis the ICSD database was used.

The electrical resistance was measured using an in-house-built high-throughput test stand (HTTS), which measures the resistance automated by the four-point-probe (4PP). The resistance measurements were performed at room temperature. The resistivity was calculated from the measured resistance values and the thin-film thickness, which was calculated from the deposition rates.

The surface morphology was examined by using the above-mentioned SEM (JEOL JSM-7200F). Additionally, the surface morphology was recorded by AFM using a Bruker FastScan scanning probe microscope in PeakForce Tapping mode with ScanAsyst using a ScanAsyst Air probe (spring constant about 0.4 N m$^{-1}$).

2.3 Electrochemical Measurements

The evaluation of the OER electrocatalytic activity of the 342 MAs was conducted using a Sensolytics Scanning Droplet Cell (SDC) system integrated with an Autolab PGSTAT204 potentiostat (Metrohm Autolab B.V., Netherlands) operated via Sensolytic SDC (version 1.0.32) and Nova (version 2.1.4) softwares. In a SDC measurement, a mobile cell is positioned on a MA, and the contacted MA served as working electrode, while a Pt wire and a Ag|AgCl|3M KCl were used as a counter and reference electrode, respectively. The measurements were conducted in 0.1 M KOH electrolyte, with a pH of 12.8, which was automatically exchanged before performing the electrochemical measurements on each MA. Initially, the open circuit potential (OCP) was measured during 60 seconds and after, electrochemical impedance spectroscopy (EIS) was recorded in the frequency range of 60 kHz to 1 kHz at the OCP with a 10 mV AC amplitude followed by a linear sweep voltammogram (LSV) that was recorded between 0 and 1 V vs. Ag|AgCl|3M KCl using a 100 mV s$^{-1}$ scan rate. As additional criteria, a cutoff current of 40 µA was used to stop the LSV recording to avoid the blockage of the cell due to the bubbles generated during the OER.

The data recorded with the SDC in each MA were processed using a MATLAB script, using the X, Y positions, the EIS data (frequency, $Z_{Re}$, $Z_{Im}$) and the voltammogram data (potential, current). The solution resistance ($R_s$) was calculated for each MA using the EIS data, taking the value of $Z_{Re}$ at the minimum positive $Z_{Im}$ (see Fig. S1). The conversion of the potential recorded vs. the reference electrode to the potential on the RHE scale was done using the following formula:

$E_{RHE}[V] = E_{Ag|AgCl|3M\ KCl} + 0.210 + 0.059 \times pH - iR_s$

where $E_{Ag/AgCl/3M\ KCl}$ is the applied potential versus the Ag|AgCl|3 M KCl reference electrode, 0.210 V is the standard potential of the Ag|AgCl|3 M KCl solution at 25 °C, while 0.059 is the result of $(RT)\cdot(nF)^{-1}$, where R is the universal gas constant, T is the temperature (at 25 °C), n is the number of electrons transferred during the reaction and is equal to 1, F is the Faraday constant and i is the recorded current.

The current normalization was done to the footprint left by the SDC on each MA (Fig. S2), after electrochemical testing. The corresponding areas were calculated by processing the digital photo of the MLs after electrochemical measurements using ImageJ (see Fig. S3). [24] First the image was quantized to 8-bit grayscale color, and then it was binarized using the built-in function Auto Local Threshold which highlighted the KOH footprints and other features with relatively lighter contrast. The footprints were selected among the highlighted features using the function Analyze Particles which identifies groups of connected pixels based on the circularity and size criteria, the selection was cross-checked to ensure that all footprints were accurately included. As output of the process a csv file with the X,Y positions and the area was obtained. The electrochemical data and footprint areas were assigned to the corresponding MA numbers, considering that the numbering is increasing from left to right (lower to higher X) and from bottom to top (lower to higher Y).

The Auto Local Threshold tool computes the threshold for binarizing each pixel according to their intensity within a window of radius (in pixels) around it. Niblack thresholding method was used. [25] In addition to the radius the Niblack thresholding method requires the k and c parameters:

    pixel = ( pixel <  mean + k * standard_deviation - c) ?  if yes, the object is background

k value, c value and radius were -0.2 and 0 and 15 (15 pixels ~ 0.5 mm) respectively.

Besides, the Analyze Particle tool selects the features from the segmented image according to the specified size and circularity ranges which were set as [0.001,0.05] $cm^2$ and [0.2-1] respectively.

## 3.  Results and Discussion

The phase analysis based on the XRD-results shows that on the ML there are three phase areas, see Fig. 2a: In the La-rich region of the ML a mixture of the phases $La_2O_3$, perovskite and $La(OH)_3$ was observed. In this compositional area the thin films are hydrated during air-exposure after the deposition, therefore also $La(OH)_3$ was found in the XRD-pattern. The exemplary XRD-pattern of this phase region in Fig. 2b shows that the La-rich oxide phases are predominantly present here, as the perovskite peaks are only present with lower intensity. The O-content in this three-phase region was measured to be around 73 at.-% using XPS of two exemplary MAs. In the area with a La-content of around 50 at.-% single-phase perovskite forms. A striking feature of the XRD-pattern is the enhanced intensity of the (102) and (204) reflections relative to the (202) reflection, indicating a textured thin-film growth (see Fig. 2b). In the Co-rich area, a two-phase region of perovskite and spinel was observed. This region contains the MAs with the highest catalytic activity, especially in the Ni-rich region of this phase-constitution area (Fig. 2c). Comparison of the XRD-pattern reveals that the MAs with higher Ni-content exhibit broader and less intense peaks, which may indicate reduced crystallinity arising from the presence of multiple spinel phases or increased structural disorder

(see Fig 2b). XPS-measurements of four exemplary MAs show that the O-content in the single-phase perovskite and the two-phase regions is consistently around 60 at.-%. In general, the phase constitution in the La-Co-Ni-O library changes primarily along the compositional gradient between La and Co, while Ni appears to play only a minor role in phase formation.

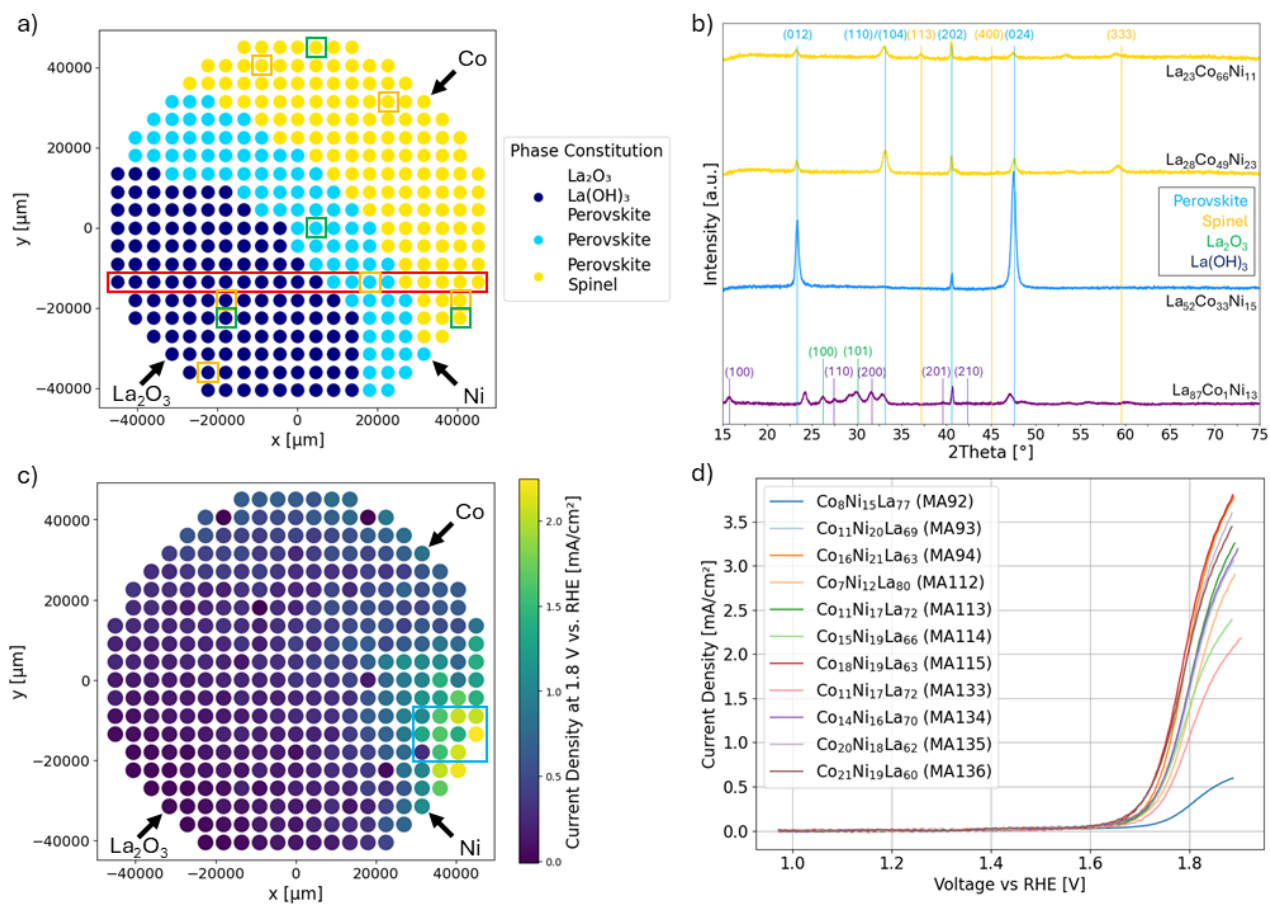

Fig. 2: a) Visualization of the phase constitution distribution over the La-Co-Ni-O thin film materials library. The location of the sputter sources with respect to the wafer are indicated. MAs marked in red are further investigated in Fig. 4, MAs marked in orange were selected for XPS analysis; b) exemplary XRD patterns for selected MAs of each phase-constitution area (marked green in Fig. 2a). Peaks are indicated using ICSD data: $La_2O_3$ ICSD-7795, $La(OH)_3$ ICSD-31584, Perovskite ($LaCoO_3$) ICSD-17668, Spinel ($Co_3O_4$) ICSD-36256.; c) color-coded OER activity map showing the current density at 1.8 V vs. RHE; d) LSVs recorded on the selected MAs (marked with a blue rectangle in Fig. 2c) containing the MAs with highest measured current density at 1.8 V vs. RHE.

The electrocatalytic activity of the 342 MAs present on the ML is presented in Fig. 2c. The highest activity was observed for few MAs in the perovskite/spinel region of the ML. The highest current density (2.24 mA/cm$^2$) was measured at a composition around $La_{11}Co_{20}Ni_9O_{60}$. However, for the same phase constitution and approximately the same Ni-content quite different OER current densities were observed. This can be better observed in Fig. 2d that shows the linear sweep voltammograms (LSVs) of the MAs on which the highest current density was recorded. Interestingly, in this comparison, we observed some deviations, which we investigated further to confirm the catalytic trends and exclude outliers. For example, MA 92 (Fig. 2c, lower left MA in corner of the blue box), located in the region of generally higher catalytic activity (as marked in Fig. 2c), exhibited a significantly reduced current density compared to the neighboring MAs. In this case, also a larger footprint left by the SDC head on the MA was observed: We assume that during retraction of the SDC head, a spread of the electrolyte may occur leading to an overestimation of the actual measured area, and thus to the apparent lower electrocatalytic activity. Therefore, these measurements can be excluded, as their footprint differs significantly from the overall area of the other footprints (see Supporting Information, Fig. S2, S3), since they seem to be a consequence of the high-throughput screening and automatic data processing. The data obtained from high-throughput measurements are further analyzed in several correlation plots. Fig. 3 shows the OER current density in dependence of the Ni-content for the three phase constitution areas of the ML. The three-phase area shows almost no activity, which is correlated to the high resistivity of the material. The single-phase perovskite MAs shows low activity with a few exceptions in the Ni-rich region. This hints at an influence of the Ni-content on the current density in the single-phase perovskite. This trend is consistent with previous reports showing performance improvements for OER in La-Co-Ni perovskites with increased surface presence of Ni. [26–28] As mentioned, the highest current density values were observed in the spinel/perovskite area. The current density increases with increasing Ni-content in this two-phase area. The conductivity of the thin films seems to have a minor influence on catalytic activity, since the MAs with highest measured current density are not exclusively characterized by a remarkably low resistivity. However, MAs with high resistivity over 100 Ω cm show almost no catalytic activity, i.e., a certain conductivity is necessary for electrocatalytic activity. Fig. 4 shows the current density in dependence of the Co-content. As for the Ni-content in Fig. 3, a dependence of the current density on the phase constitution is observed. Furthermore, the influence of the Co-content on the phase constitution is illustrated, as the different phase constitutions are clearly separated along the axis for Co-content. Moreover, an influence of composition on the resistivity of the material is visible, since almost all MAs with resistivities below 0.5 Ω cm are found between 30 - 50 at.-% Co. This area with lowest resistivity is in the region of single-phase perovskite and spinel/perovskite multiphase. For the two-phase region the lower resistivity values are near the single-phase region, in which a higher proportion of perovskite should be present (see Fig. 2a and Fig. S4). For the two-phase region, between approximately 40 - 60 at.-% Co (normalized to the metals in the thin film) the values for the current density are scattered, which indicates a minor influence of the Co-content on activity. For higher Co-contents the activity could be lower, because there is less Ni in the material. This correlation is shown in Fig. 5: For the perovskite/spinel two-phase region, the MAs with highest measured current densities are located in the Ni-rich region.

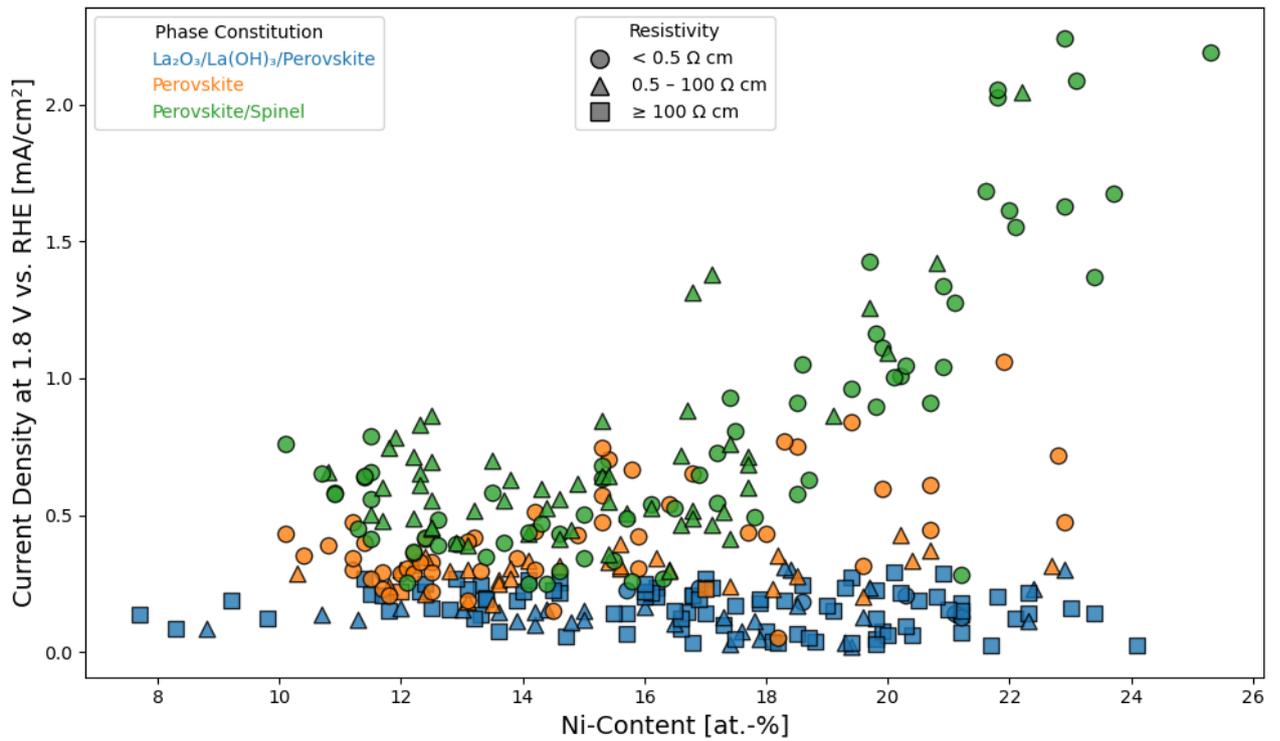

Fig. 3: OER current density values at a potential of 1.8 V vs. RHE plotted in dependence of the Ni-content. The phase constitution of the MAs is color-coded, different areas for the electrical resistivity are coded in marker form. Highest values are obtained for perovskite/spinel two-phase materials with a high Ni content.

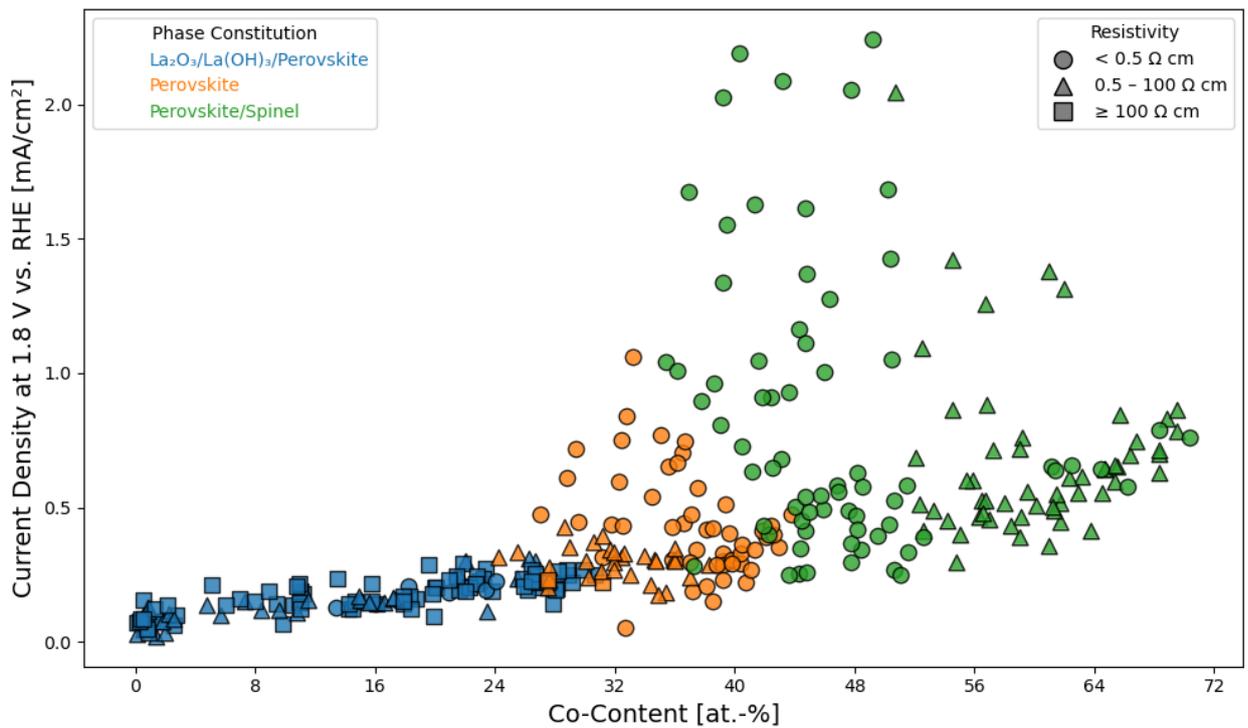

Fig. 4: OER current density values at a potential of 1.8 V vs. RHE plotted in dependence of the Co-content. The phase constitution of the MAs is color-coded, different areas for the resistivity are coded in marker form. Highest values are obtained for perovskite/spinel two-phase materials with wide scattering for Co-contents between 40 at.-% and 60 at.-%.

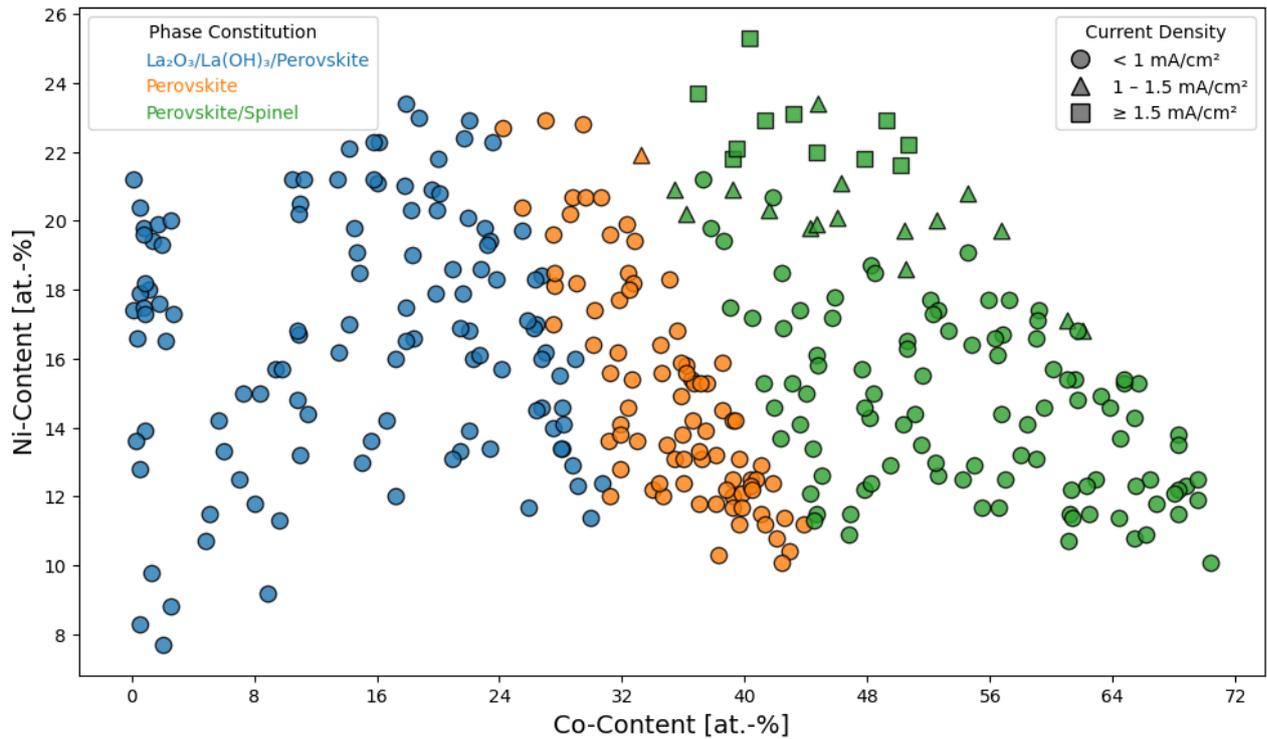

Fig. 5: Ni-content plotted over Co-content. The phase constitution of the MAs is color-coded, different areas for current densities are coded in marker form. MAs showing highest catalytic activity can mostly be found in the two-phase region with high Ni-content.

Fig. 6 is used to infer correlations between OER current density and electrical resistivity. For this, MAs along a line across all three regions with different phase constitutions were selected (see Fig. 2a, MAs in red box). A comparison between current density and resistivity shows that higher current densities were measured for films with lower electrical resistivity (see Fig. 6). A dependence is also seen with respect to the phase constitution, which has already been observed in Fig. 3 and 4. It is not clear if the increase of current density is caused by the change in crystal structure or by the lower resistivity or whether both have an influence. The increase of the current density within the single-phase perovskite region and two-phase region correlates with an increasing Ni-content, again.

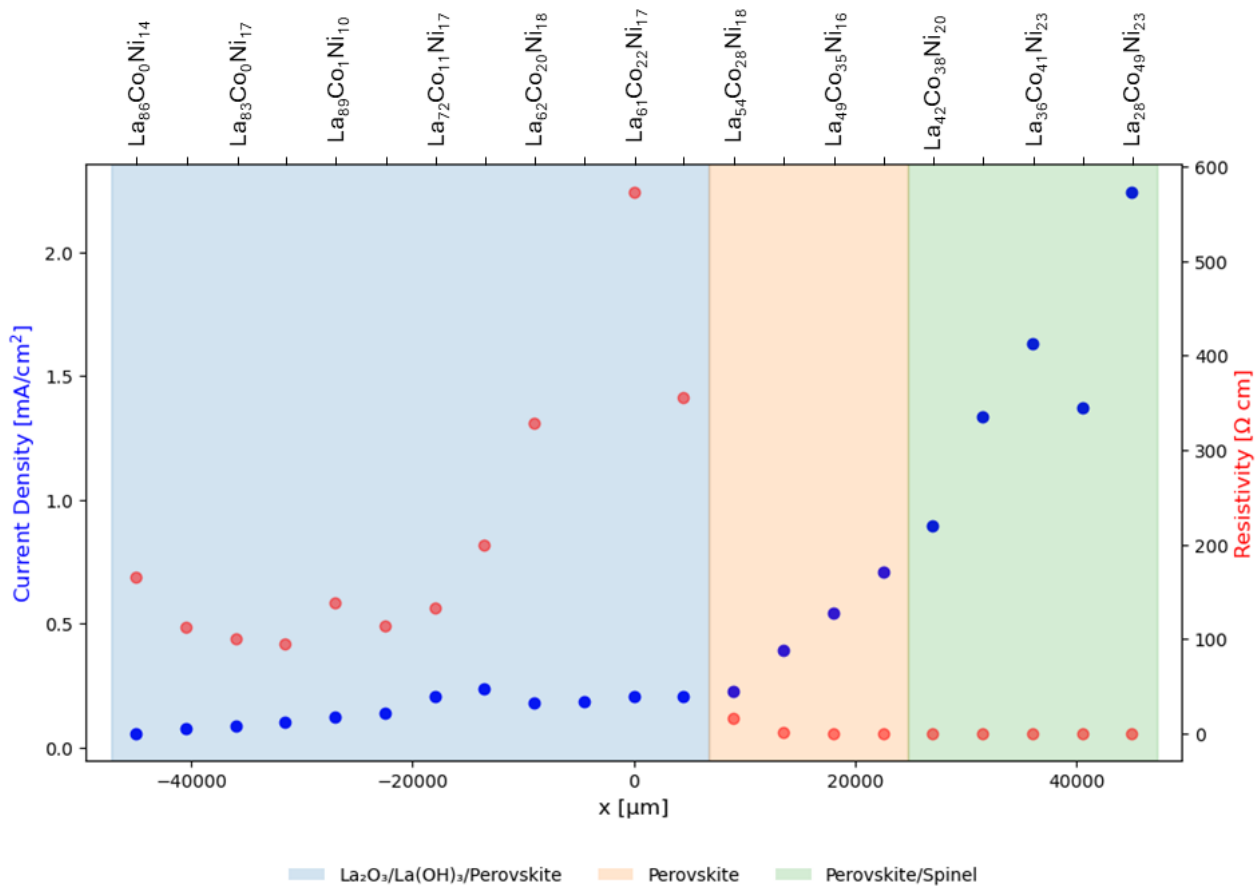

Fig. 6: OER current density values (blue) and electrical resistivity values (red) plotted in dependence of location on the ML crossing the three phase-constitution areas. The chemical composition is indicated normalized to the metals. The current density increases in the perovskite and two-phase region toward the Ni-rich area. The resistivity drops abruptly in the perovskite area compared to the three-phase region and stays at low values for the two-phase region comprising perovskite. The selected MAs are marked red in Fig. 2a.

An influencing factor on the observed activity trends could be different surface morphologies and surface roughness as they are related to the electrochemical surface area. Fig. 7 compares the surface morphology in the single-phase perovskite to the perovskite/spinel two-phase region. In the perovskite region a nanocrystalline microstructure was observed with a relatively smooth surface, as the height variations are relatively low with a maximum of 60 nm. The surface microstructure is homogeneous without noticeable segregation for the single-phase perovskite. In the two-phase region, there is a similar surface microstructure, but there are also additional crystallites grown on the surface, which increase the roughness and thus also the electrochemical surface area. The height differences are significantly larger here and reach up to over 300 nm. With increasing Ni-content an increasing number of these crystallites with a uniform distribution on the surface was observed. Two differently shaped types of crystallites were observed: Spherical-like with diameters of around 300 nm and smaller triangular-like with edge lengths of mostly around 100 - 200 nm.

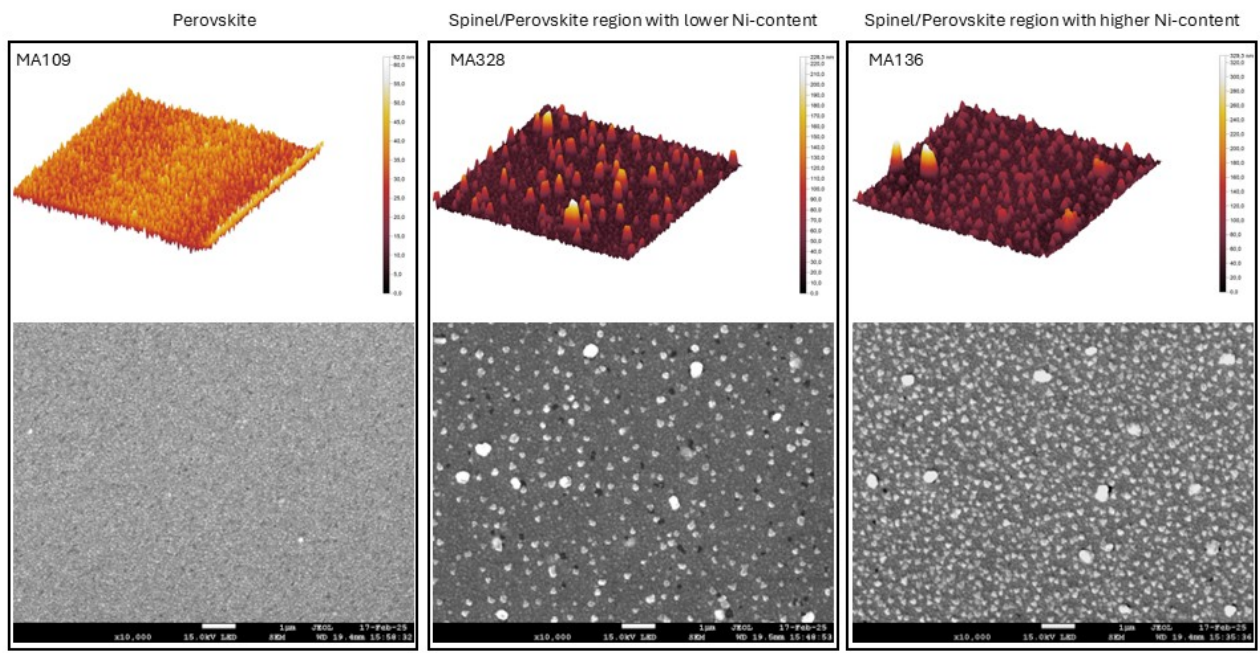

Fig. 7: Comparison of surface morphology between single-phase perovskite and spinel/perovskite two-phase region with different Ni-contents. The AFM and SEM results show a nanocrystalline surface microstructure for the perovskite and two-phase regions. Larger crystallites can be found on the surface in the two-phase region, which increase in number with increasing Ni-content. The AFM images display a scanned area of 5 µm x 5 µm. The scale bar on the SEM images (lower row) are 1 µm.

The surface morphology of the two-phase region is shown at higher magnification along with chemical analysis in Fig. 8. EDX analysis shows that the crystallites, both the spherical and triangular ones, on the surface are Co- and Ni-rich. In the underlying material a generally homogeneous elemental distribution was observed. The triangular-like shape of the crystallites, typical for (111)-surface of spinels, indicate the presence of Co-Ni-spinels on the Ni-rich surface. [29–31] This hypothesis is further supported by the observed correlation between the increased surface crystallite formation and the higher Ni-content, as confirmed by EDX, which also indicates that these surface features are associated with a Ni-containing phase. However, this assumption cannot be conclusively confirmed on the current data and requires further in-depth structural and compositional characterization which is beyond the scope of this work. The presence of Co-Ni-spinels in the Ni-rich region may contribute to the highest activity for OER observed in that region. Several works have pointed out that a combination of different Ni and Co cations forms Ni-Co oxyhydroxides, a structure in which the formation of highly active Co species is facilitated. [32,33] Likewise, the presence of Ni may itself contribute to the higher observed OER activity, as Ni-enriched surfaces in Co–Ni spinels have been reported to be more active in Co-Ni spinels [34] and La-Co-Ni perovskites [27].

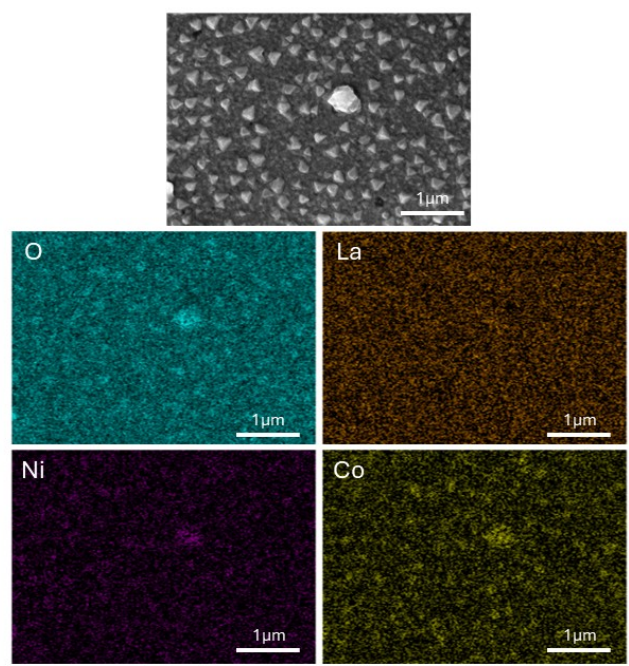

Fig. 8: SEM image of the surface microstructure with higher magnification and element distribution in the Ni-rich area of the two-phase region measured via EDX. The crystallites on the surface contain Co and Ni and are mostly triangular-shaped, which suggests that these could be spinels. The overall chemical composition in this measurement area normalized to the metals of the thin film was determined to be $La_{25}Co_{53}Ni_{22}$ by EDX.

Apart from the reasons already discussed for the enhanced catalytic performance in the two-phase region, the coexistence of distinct phases itself may contribute beneficially to catalytic activity. Heterostructures can be advantageous for the properties of a material due to synergy effects, for example stress effects at the interfaces or electronic interactions between the phases. [35] Wang et al., for instance, already reported on improved catalytic activity through favored activation of lattice oxygen at spinel/perovskite interfaces. [36] These insights point toward the potential relevance of similar interfacial phenomena in the present system.

## 4. Conclusions

Results from combinatorial synthesis and high-throughput characterization of a La-Co-Ni-O ML were presented. Three regions with different phase constitutions were identified: In the La-rich area a three-phase region of $La_2O_3$, $La(OH)_3$ and perovskite forms. In the Co-rich area, a two-phase region of spinel and perovskite forms. Between these multi-phase regions, a region of single-phase perovskite was identified. The ML shows the highest electrocatalytic activity for OER in the perovskite/spinel two-phase region and activity is increasing with higher Ni-content. The Co-content on the other hand influences the phase constitution. Correlating the measured properties showed that lower electrical resistivity of a material could be beneficial for its catalytic performance. However, it cannot be explained conclusively if the increased electrocatalytic activity in this case is caused by the lower electrical resistivity or just by the phase constitution or both. Analysis of the surface microstructure shows crystallites with high contents of Co and Ni grown on the sample in this two-phase region, which could be an influencing factor for the higher catalytic activity due to an increased surface area. In summary, we found that multiphase, Ni-rich films with a composition of $La_{11}Co_{20}Ni_9O_{60}$ are most promising in the La-Co-Ni-O system. Increasing the Ni-content and adding additional cations towards high-entropy oxides might enhance the catalytic performance further.

**Data Availability**

The dataset for this publication, including the compositional dataset, electrochemical dataset, the diffractograms, microscopy images and resistance measurements, is available at Zenodo under https://doi.org/10.5281/zenodo.15754300.
Additionally, the data analyzed in this work can be found in the used research data management system [37] under the following sample IDs: 0010669, 0010671.

**Acknowledgments**

This work was funded by the Deutsche Forschungsgemeinschaft (DFG, German Research Foundation), project number 388390466-TRR 247, projects C04 and A02. The Center for Interface-Dominated High Performance Materials (ZGH, Ruhr University Bochum) is acknowledged for SEM, EDX, XRD and AFM measurements.

**Keywords:** perovskites, oxygen evolution reaction, transition metal oxides, reactive magnetron sputter deposition, high-throughput experiments

## Supporting Information

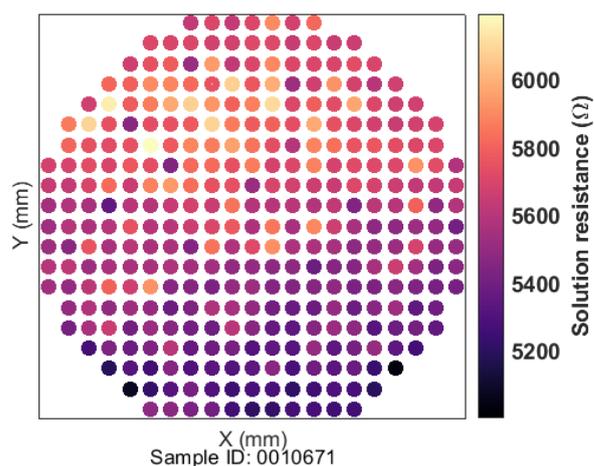

Fig. S1: Visualization of solution resistance during the electrochemical measurements of the La-Co-Ni-O thin film materials library.

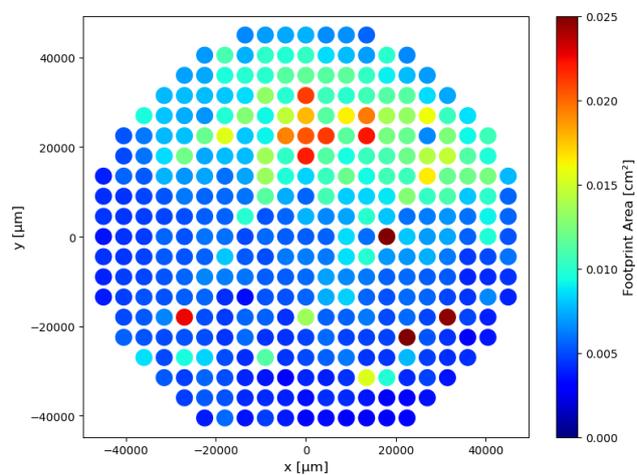

Fig. S2: Footprint area of the SDC-tip on the different MAs after OER measurements.

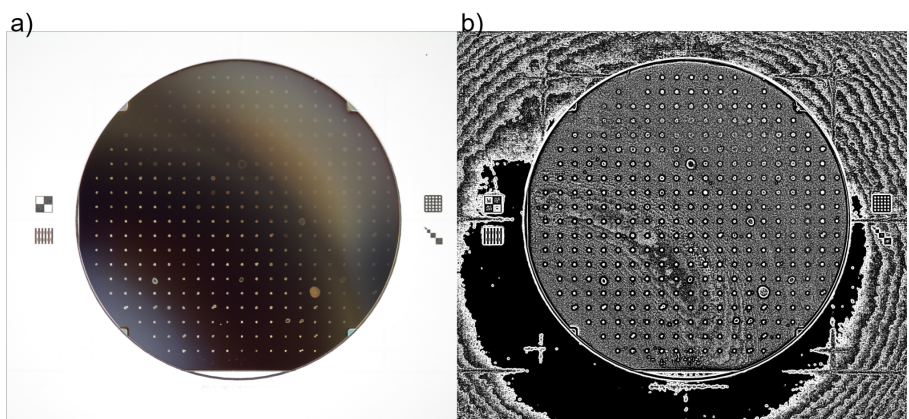

Fig. S3: a) Digital photo taken after electrochemical measurements. b) Image quantized to 8-bit and binarized, the footprints are highlighted in white.

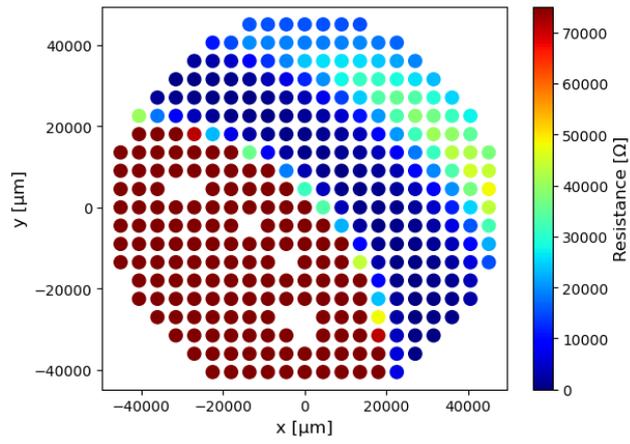
Fig. S4: Measured electrical resistance by 4PP-HTTS